\documentclass[reprint, twocolumn,superscriptaddress, showpacs,aps,prl]{revtex4-1}

\usepackage{amsmath} 
\usepackage{amssymb} 
 \usepackage{amsfonts}

\usepackage{graphicx} 

\usepackage{array}
\usepackage{multirow}
\usepackage{color}
\usepackage{transparent}
\usepackage{float}

\newcommand{\vect}[1]{\boldsymbol{\mathbf{#1}}}

\begin{document}

\bibliographystyle{apsrev4-1}

\title{Resonance Reaction in Diffusion-Influenced Bimolecular Reactions} 

\author{Jakob J. Kolb}
\affiliation{Institut f{\"u}r Physik, Humboldt-Universit{\"a}t zu Berlin, 12489 Berlin, Germany}
\author{Stefano Angioletti-Uberti}
\affiliation{International Center for Soft Matter Research, Beijing University of Chemical Technology, Beijing 100029, China}
\author{Joachim Dzubiella}
\email{joachim.dzubiella@helmholtz-berlin.de}
\affiliation{Institut f{\"u}r Physik, Humboldt-Universit{\"a}t zu Berlin, 12489 Berlin, Germany}
\affiliation{Institut f{\"u}r Weiche Materie und Funktionale Materialen, Helmholtz-Zentrum Berlin f{\"u}r Materialien und Energie, 14109 Berlin, Germany}


\date{\today}

\begin{abstract}

We investigate the influence of a stochastically fluctuating step-barrier potential on bimolecular reaction rates by exact analytical 
theory and stochastic simulations. We demonstrate that the system exhibits a new 'resonant reaction' behavior with rate enhancement 
if an appropriately defined fluctuation decay length is of the order of the system size. Importantly, we find 
that in the proximity of resonance the standard reciprocal additivity law for diffusion and surface reaction rates is violated due to the 
dynamical coupling of multiple kinetic processes. Together,
these findings may have important repercussions on the correct interpretation of various kinetic reaction problems in complex systems, 
as, e.g., in biomolecular association or catalysis.

\end{abstract}

\maketitle

Bimolecular reactions constitute  key processes for function in physical chemistry and biology~\cite{Calef1983, berg1985diffusion}. The first step in such a reaction is the diffusive particle approach which gives rise to an intrinsic diffusion rate $k$. The second step involves a chemical reaction once particles are close to contact, described by a surface rate $k_{\rm surf}$.  The mean rate of the total reaction is provided by the standard 
reciprocal additivity law  
\begin{eqnarray}
k^{-1}_{\rm tot} = k^{-1} + k^{-1}_{\rm surf}.
\label{KS_nonideal}
\end{eqnarray}
In a simple two-body picture, typically the famous Smoluchowski-Debye expression $k_{\rm S}$ for the diffusion-controlled rate $k$ is employed, describing the diffusive encounter rate of a particle with diffusion constant $D$ to reach the second particle modeled as a spherical sink with effective radius $R_{\rm s}$. If the diffusion proceeds across a static energy landscape $U(r)$, the
final expression is~\cite{Calef1983, berg1985diffusion, Smoluchowski1917a,Debye1942}
\begin{eqnarray}
k_{\rm S}^{-1} = \int_{R_{\rm s}}^\infty  {\rm d}r \frac{\exp[\beta U(r)]}{4\pi D r^2}. 
\label{S_rate}
\end{eqnarray}
However, in  complex systems that exhibit multiple degrees of freedom, the effective potential energy $U(r)$ along a convenient reaction coordinate $r$ may thermally fluctuate in space and time between multiple states~\cite{Calef1983, berg1985diffusion,kang,zwanzig}. Relevant examples can be found in the binding  of ligands to conformationally-gated proteins \cite{Szabo1982, agmon, srajer, zhou, greives} or weakly hydrophobic 
pockets \cite{Setny2013, Mondal2013},  association kinetics of biomolecules with fluctuating charges \cite{kirkwood,mikael}, polymer translocation \cite{polymertranslocation} or folding~\cite{wonkyu}, and catalytic reactions in stimuli-responsive nanoreactors \cite{Wu2012a, nanoreactor}.  

In those cases, one can expect significant alteration of total reaction rates originating from fluctuations of the energy landscape, 
as indicated by the very related, but 'inverse' problem of the activated escape over fluctuating potential barriers~\cite{Doering1992, Zurcher1993, Pechukas1994, Reimann1995, Reimann1995a, Reimann1997, mantegna, schmitt}. Here,  a fascinating resonance phenomenon,  
called 'resonant activation', with rate enhancement at crtitcal fluctuation time scales was observed.  That this phenomenon falls not into the framework of stochastic resonance has been nicely discussed in the paper by Schmitt {\it et al.}~\cite{schmitt}.  However, to the best of our knowledge, and somewhat 
surprisingly given the wealth of literature on resonant activation, the consequence of barrier fluctuations on diffusion-influenced 
reactions has  not been explored, yet. 

In this communication, we close this gap by studying the problem of diffusion-influenced reaction rates in the presence of a spherically-symmetric step-barrier potential fluctuating between multiple states within the classical, spherical Smoluchowski-Debye setup. 
This model system, while still a valid prototype and approximation of many important realistic scenarios, directly applies to the geometry of so-called yolk-shell nanoreactors where a central catalyst (sink) is embedded within a hydrogel shell~\cite{Wu2012a, nanoreactor}. Near its critical solution temperature the polymer shell strongly fluctuates [27] and an unexplained dip appears superimposed on the temperature dependence of the rate predicted by standard theory~\cite{Wu2012a}.  Here, we demonstrate by both exact analytical theory 
and stochastic simulations that the phenomenon of resonance with rate enhancement can indeed be observed in diffusion-influenced reactions 
if  the time scale of the barrier fluctuations couple to those of the diffusion-reaction process. Secondly, we show that in the proximity of the 'resonance reaction' the standard (and exact for non-fluctuating barriers) additivity law for diffusion and surface  reactions eq.~(\ref{KS_nonideal}) is violated due to the multiple dynamical coupling of time scales. Together,
these findings have important repercussions on the correct interpretation of various kinetic reaction problems in complex, fluctuating systems~\cite{Szabo1982, agmon, srajer, zhou, greives, Setny2013, Mondal2013, kirkwood,mikael, polymertranslocation, wonkyu, Wu2012a, nanoreactor}.

\begin{figure}[t]
\begin{center}
	\includegraphics[width = 0.35 \textwidth]{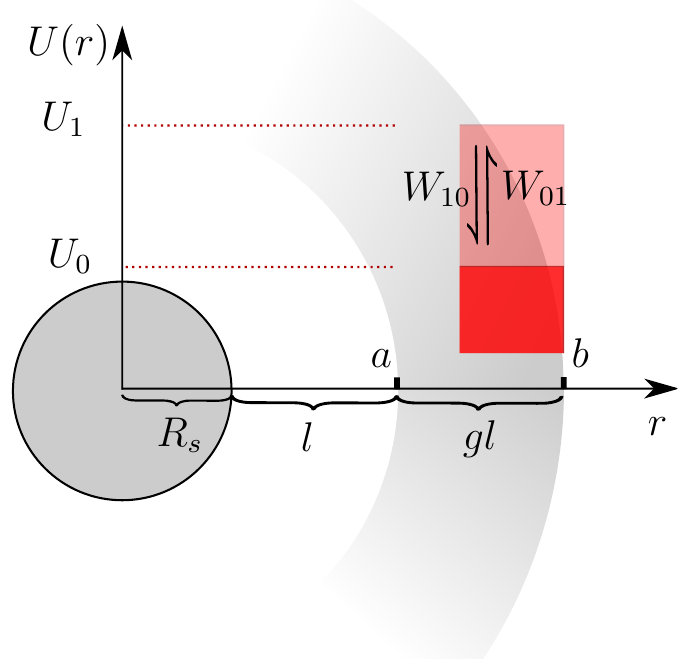}
    \caption{Sketch of our model system consisting of a spherical sink particle (grey sphere) and a fluctuating step-barrier (red). 
    The sink radius is $R_{\rm s}$, while the barrier is positioned between radial distances $a$ and $b$. The gap between the sink and the  barrier is 
    thus $l=a-R_{\rm s}$.  The scaled barrier width we then define as $gl=b-a$. In this illustration, the barrier fluctuates between two states (dark and light red) 
    with barrier heights $U_0$ and $U_1$ and transition rates $W_{01}, W_{10}$, respectively.}
\label{fig0}
\end{center}
\end{figure}

Our minimalistic model is illustrated in Fig.~\ref{fig0}. 
As in the classical Smoluchowski-Debye picture for diffusion-controlled reactions~\cite{Calef1983, berg1985diffusion} over static potentials,  the diffusional approach of ideal reactants over an energy landscape towards a central spherical sink with radius $R_{\rm s}$ is considered.  
We set $R_{\rm s}=1$  as our unit length scale in the remainder of the paper.  As energy landscape, we consider a step-barrier potential, defined by the piece-wise function $U_n(r) = U_n \left[\Theta(r-a)-\Theta(r-b)\right]$,  where $U_n$ is the barrier height of state $n$ of $N$ possible states, $l=a-R_{\rm s}$ is its radial distance to the sink surface, and we define the barrier width as $gl=b-a$, $g$ being the ratio 
between barrier width and the length $l$.   In this way, as long as $g\approx 1$, $l$ is a convenient measure of the system size. 
The spatial intervals in $r$ with constant potential, namely $R_{\rm s} \le r < a$, $\; a \le r <  b$, and $r>b$ will be referred to as (I), (II), and (III) in the following, respectively.  Note that in Fig.~\ref{fig0} only two states are exemplified while our mathematical approach is completely general for multiple states.

We now assume that the barrier height switches stochastically between the $N$ states according to a discrete time reversible 
Markov process $\eta(t)$. \textcolor{black}{If we further consider that this process is not influenced by the reactant, or in other words that the probability for the external potential to be in a specific state does not depend on the presence of the reactant, the evolution of 
the coordinate of a single reactant} follows the stochastic differential equation (SDE) 
\begin{equation}
    \frac{{\rm d} \vec{r}}{{\rm d} t} = \vec{\nabla}\frac{1}{\gamma}f(r)\eta(t) + \sqrt{2D}\vec{\varepsilon}(t), 
    \label{SDE}
\end{equation}
where $\varepsilon(t)$ is white Gaussian noise with time correlation $\left< \varepsilon(t) \varepsilon(t') \right> = \delta(t-t')$, and $\eta(t) \in [U_0,\cdots U_{N-1}]$ and $f(r) = \Theta(r-a)-\Theta(r-b)$ define the height and shape of the potential barrier. The friction constant $\gamma$ sets our time scale and is related to the reactant self-diffusion constant $D$ through the Einstein relation $\gamma = k_{\rm B}T/D$, where $k_{\rm B} T$ is the thermal energy, our natural energy scale.  
\textcolor{black}{As implicit in all works using the classical Smoluchowski-Debye picture of diffusion on an energy landscape, decoupling the state of the external environment from the reactant field as we also do here is an {\it approximation}. This approximation should be a valid assumption for weakly interacting and dilute reactants, but is likely to break down for strongly correlated systems.}

An equivalent description of the problem can be given in terms of a combined reaction-diffusion equation for the particle density function $\rho_n(\vec{r},t)$ 
of the discrete variable $n=0,..,N-1$ of the potential and the continuous variable $\vec{r}$ of the overdamped particles, via
\begin{equation}
    \frac{\partial}{\partial t}\vect{\rho}(\vec{r},t) = \left\{ \mathbb{F} + \mathbb{W} \right\} \vect{\rho}(\vec{r},t),
    \label{mfpe}
\end{equation}
where $\mathbb{F}$ is the Fokker-Planck operator
\textcolor{black}{
\begin{equation}
\mathbb{F} = {\rm  diag}\left[\vec{\nabla} \frac{U_n}{\gamma} \left( \delta(r-a) - \delta(r-b) \right)\hat{e}_r + D\vec{\nabla}^{2} \right]
    \label{FPO}
\end{equation}
with $\hat{e}_r$ being the unit vector in radial direction}, and $\mathbb{W}$ is the transition rate matrix of the Markov process for the barrier switching. $\vect{\rho}(\vec{r},t)=(\rho_0(\vec{r},t),\cdots,\rho_{N-1}(\vec{r},t))^{T}$ denotes the vector of particle density functions related to each state of the potential barrier. \textcolor{black}{Since the underlying Markov process of $\mathbb{W}$ is time reversible the transition rate matrix satisfies detailed balance, i.e. $ W_{mn} \exp(-\beta F_n) = W_{nm} \exp(-\beta F_m) $, where $W_{mn}$ is the switching rate from state $n$ to state $m$ of the external potential, and $F_{n(m)}$ is the underlying free-energy of the system/environment determining the external potential in the $n(m)$ state. This free-energy should not be confused with the potential energy {\it felt by the reactant in the $n(m)$ state}, previously labeled $U_n$} This also implies that the particle density vector at infinite distance, where any effect due to the potential is lost, is simply equal to the \textcolor{black}{{\it constant} equilibrium (bulk) vector $\vect{\rho}^{(eq)}$ of $\mathbb{W}$}.

With these prerequisites it is now possible to find a similarity transform $\mathbb{T}_{ij}=[\rho_n^{(eq)}]^{1/2}\delta_{i,j}$ such that the resulting $\mathbb{T}^{-1}\mathbb{W}\mathbb{T} = \mathbb{S}$ is symmetric \cite{Oppenheim1977}. This symmetric matrix can then be diagonalized by an orthogonal transformation $\mathbb{D}$ resulting in $\mathbb{D}^{\dagger}\mathbb{S}\mathbb{D} = - {\rm diag}[\lambda_n]$. It can be shown \cite{VanKampen1992} that $\lambda_{n>0}>0$ and $\lambda_0=0$ with corresponding eigenvector $\mathbb{D}_{0,n}=[\rho_n^{(eq)}]^{1/2}$.   Therefore we can give a steady-state solution $\vect{\rho}(\vec{r}) = \mathbb{T}\mathbb{D}\tilde{\vect{\rho}}(\vec{r})$ to eq.~\eqref{mfpe} in terms of eigenfunctions of $\mathbb{W}$ via
\begin{align}
    \label{solution}
    \tilde{\rho}_{0}^{(j)}(r) &= c_{0,1}^{(j)} + c_{0,2}^{(j)} \frac{1}{r} \\
    \tilde{\rho}_{n \ne 0}^{(j)}(r) &= c_{n,1}^{(j)}\frac{1}{r} \exp\left[-r\sqrt{\frac{\lambda_n}{D}}\right] + c_{n,2}^{(j)}\frac{1}{r} \exp\left[r\sqrt{\frac{\lambda_n}{D}}\right]  \nonumber
\end{align}
separately for the regions $j=$~(I), (II), and (III), exploiting the fact that the Fokker-Planck operator $\mathbb{F}$ is invariant under the transformations 
$\mathbb{T}$ and $\mathbb{D}$ for $r\ne a, b$. The coefficients $c^{(j)}_{n,k}$ have to be obtained from boundary (density and flux) matching conditions at $r=a,b$, see the Supplemental Material (SM)~\cite{supp}.  
From the exact solution (\ref{solution}) it is visible that the spatial influence of the potential fluctuations decays with a certain \textit{fluctuation decay length} equal to
\begin{equation}
    r_{\rm d} = \left\{\sqrt{\frac{\lambda_m}{D}}\right\}^{-1}
    \label{decay_length}
\end{equation}
that only depends on the diffusion constant $D$ of the Brownian particles and the largest nonzero eigenvalue $\lambda_m$ 
of the transition rate matrix. In a simple two-state case, $\lambda_m$ expresses essentially the transition rate between the two states. 
Hence, the decay length describes the mean diffusive path of a particle after its disturbance by a fluctuation and thus is a measure for the spatial 
range of the action of the fluctuation.

Given the general form of the density profiles eqs.~(\ref{solution}), the diffusion-influenced reaction rate is given by
\begin{equation}
    \rho_{\infty}k = 4 \pi D R_{\rm s}^{2}\sum_n \left. \frac{\partial \rho_n^{(I)}(r)}{\partial r} \right|_{R_{\rm s}},
    \label{rate_konstant}
\end{equation}
where the density is calculated by imposing the proper boundary condition at the sink, and \textcolor{black}{$\rho_\infty = \sum_n \rho^{eq}_n$ with 
$\rho^{eq}$ being the equilibrium (bulk) vector according to $\mathbb W$}. For perfectly
adsorbing conditions $\rho_n(r=R_{\rm s})=0$ and \textcolor{black}{$\rho_n(r\rightarrow\infty)=\rho^{eq}_n$}, whereas for partially adsorbing boundaries (so-called radiative boundary conditions)
the density and its derivative at the sink are coupled through the equation:
\begin{equation}
    4 \pi D R_{\rm s}^2 \sum_n \left. \frac{\partial \rho_n^{I}(r)}{\partial r} \right|_{R_{\rm s}} = k_{\rm surf} \sum_n \left. \rho_n^{(I)}(r) \right|_{R_{\rm s}}.
    \label{nonideal_sink}
\end{equation}\\
The coefficients $c^{(k)}_{i,j}$ are calculated from the boundary and matching conditions analytically via a Mathematica~\cite{mathematica} script. 
The density profiles  and the resulting reaction rate are obtained via eqs.~(\ref{solution}) and via (\ref{rate_konstant}) or (\ref{nonideal_sink}). 

The simplest possible setup in the just developed analytical framework is that of a two state barrier that switches between 
one \emph{off} ($U_0 = 0$) and one \emph{on} ($U_1 \ne 0$) state symmetrically, i.e., the \emph{on} $\rightarrow$ \emph{off} and \emph{off} $\rightarrow$ \emph{on} rates are equal, that is, $W_{01}=W_{10}=W$, \textcolor{black}{i.e., the free energy of the environment of these two states is the same.}  We study this minimalistic two-state setup because it provides a clean basis for a detailed study of effects solely coming from the coupling of the individual time scales of 
barrier fluctuations and diffusive transport, without any complexity of having a spectrum of time scales. 
In this case, it can be easily shown that  the eigenvalues are $\lambda_0 = 0$ and $\lambda_1 = 2W$, giving a 
fluctuation decay length of $r_{\rm d} = \sqrt{{D}/{(2W)}}$.

\begin{figure}[t]
\begin{center}
\includegraphics[width= .35 \textwidth]{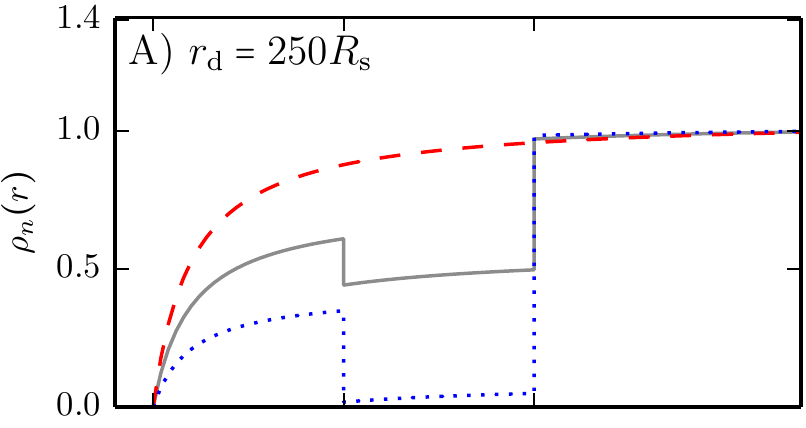}
\includegraphics[width= .35 \textwidth]{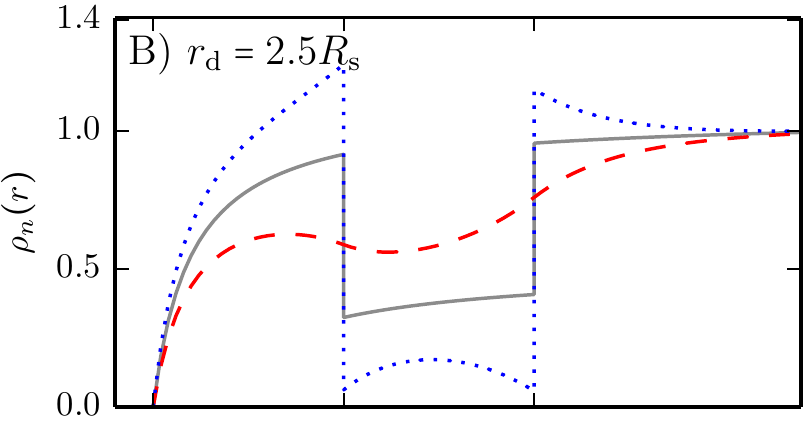}
\includegraphics[width= .35 \textwidth]{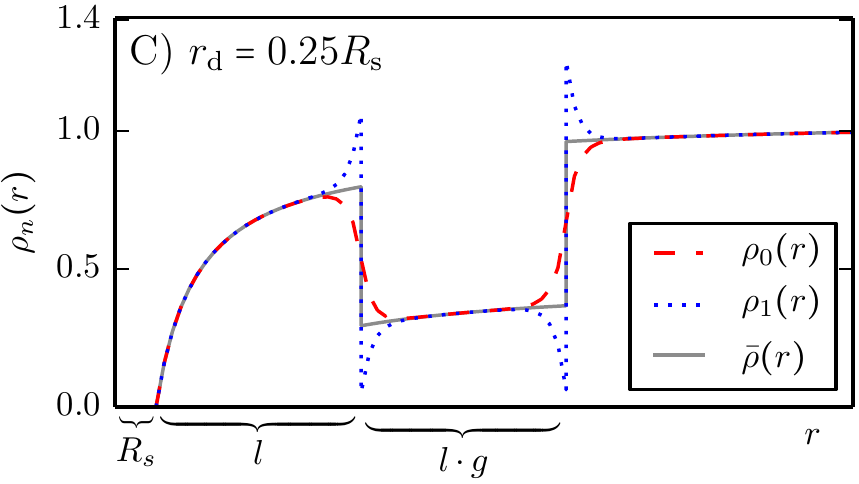}
\caption{Analytic results for steady-state density profiles $\rho_0, \rho_1$ for two states of the fluctuating repulsive barrier. Also shown is the total density profile $\bar{\rho} = (\rho_0+\rho_1)/2$. All parameters but the decay length are fixed: $a = 6 R_{\rm s}$, $b = 11 R_{\rm s}$, $l=5$, $g=1$, $U_0 = 0$, $U_1 = 3~k_{\rm B}T$. The decay length is A) $r_{\rm d} = 250$, B) $r_{\rm d}=2.5$, and C) $r_{\rm d}=0.25$.}
\label{fig1}
\end{center}
\end{figure}

We independently check our analytical treatment using numerical Brownian Dynamics (BD) simulations~\cite{andy} where the single-particle SDE eq.~\eqref{SDE} is discretized in time and then used to describe an ensemble of independent particle trajectories. Details on the analytical and numerical evaluations, \textcolor{black}{in particular the lengthy (but exact) equation for the rate in the two-state case} can be found in the SM~\cite{supp}. 

For the aforementioned two-state symmetric system, we calculate the radial steady-state density profiles $\rho_n$ resulting from the reverse transform of eq.~\eqref{solution}. The results for the density profiles for fully adsorbing boundary conditions and parameters  $U_1=3~k_{\rm B}T$, $l=5$, and $g=1$ and for three different 
transition rates, expressed by $r_{\rm d} = 250$, 2.5, and 0.25, are shown in Fig.~\ref{fig1}. We also plot the total density profile $\bar{\rho} = (\rho_0+\rho_1)/2$. 
A qualitative consideration of these results shows that for small rates (large decay length $r_{\rm d} = 250$, panel A), the profiles $\rho_n$ are close to their respective steady-state distributions~\cite{Dzubiella2005} without any switching. In this slow fluctuation limit, the total profile $\bar{\rho}$ is thus given essentially by 
the weighted sum of the respective steady-state distributions. For high rates, (small decay length $r_{\rm d} = 0.25$, panel C) the steady-state profiles are all very similar: perturbations  are on a small time scale and the profiles converge to the same limit where the reactants see an average potential barrier of height $\bar U = 1.5~k_{\rm B}T$.  These slow and fast limits are also known in escape problems over fluctuating barriers~\cite{Doering1992}. Intermediate, much more complex behavior is observed for values of the decay length comparable to the barrier dimensions ($r_{\rm d} = 2.5$, panel B).  Now the perturbations are 
significant on the system scale. In particular, note that  for this intermediate $r_{\rm d}$ the total density $\bar\rho$ between sink and barrier is higher
than for both the slow and fast limits, indicating an increase of reactants close to the sink 'pumped' by the fluctuating barrier. 

\begin{figure}[t]
\begin{center}
\includegraphics[width= .35 \textwidth]{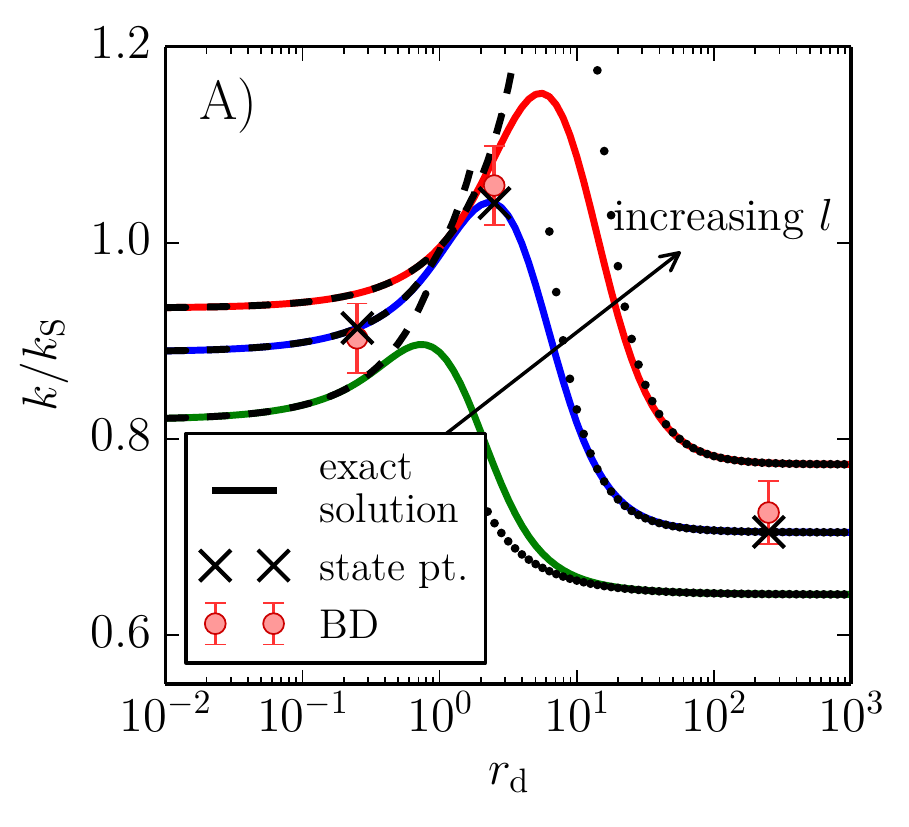}
\includegraphics[width= .35 \textwidth]{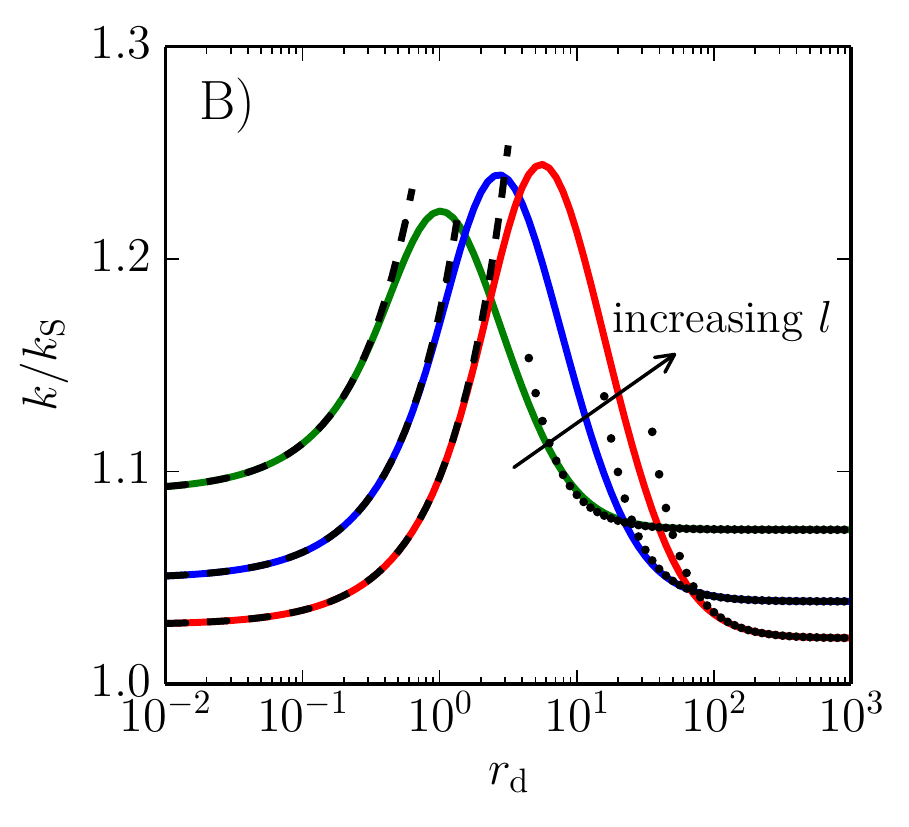}
\caption{The diffusion-controlled rate $k$ vs. decay length $r_{\rm d}$ for a repulsive (panel A, $U_1 = 3 ~k_{\rm B}T$) and an attractive 
(panel B, $U_1 = -3 ~k_{\rm B}T$) fluctuating barrier for varying overall system size $l = 2, 5, 10$. Other parameters are $U_0= 0$ and $g = 1$. 
The reaction rate is normalized to the Smoluchowski rate $k_{\rm S} = 4\pi D R_{\rm s}$ of an ideal sink without barrier, cf. eq.~(\ref{S_rate}). Simple analytical 
forms for the slow and fast limits~\cite{supp} are depicted in dashed an dotted lines, respectively. State points of the density profiles at $r_{\rm d} = 0.25$, 2.5, and 250  in Fig.~\ref{fig1} are marked  by black crosses. Numerical results from BD simulations are depicted by spherical symbols with their confidence 
intervals as error bars. }\label{fig3}
\end{center}
\end{figure}

The behavior of the density profiles directly affects the resulting reaction rates, as shown in Fig.~\ref{fig3},
where we plot the reaction rate scaled by the Smoluchowski limit eq.~(\ref{S_rate}) for $U_0=0$, $k_{\rm S}=4\pi DR_{\rm s}$, 
versus the the decay length $r_{\rm d}$. The $r_{\rm d}$ values (state points) for which we discussed the density profiles are indicated by crosses.   
In panel (A) we show results for a repulsive barrier ($U_1 = 3 ~k_{\rm B} T$), as before, for various system sizes $l$.  
In panel  (B) we now also show results for an attractive well ($U_1 = -3 ~k_{\rm B} T$).   As a striking result in all curves we 
observe that at a certain decay length comparable to the system size $r_{\rm d} \simeq 1 - 10$  the reaction rate takes a maximum value. 
The decay length at which the rate is maximized increases with the system size $l$ for both repulsive barriers and attractive wells.
(This happens as well with variation of $g$; variation of $U_1$ plays a minor role if $U_1\gg k_BT$~\cite{supp}).   
Selected numerical BD solutions for $l=5$, also plotted in  Fig.~\ref{fig3}, confirm this non-monotonic behavior.  
We note that relatively simple equations for the slow and fast limits, $r_{\rm d}\gg l $ and $r_{\rm d} \ll l$ 
can be derived analytically from the (quite involved) exact solution~\cite{supp}. They are indicated by dashed and dotted lines in Fig.~\ref{fig3}
and show a rate increase for both limits when $r_{\rm d}$ tends towards values comparable to the system size. This is an analytical 
proof that a maximum rate must occur in between.  Analogously to \emph{resonant activation}~\cite{Doering1992, Zurcher1993, Pechukas1994, Reimann1995, Reimann1995a, Reimann1997, mantegna}, we can coin this yet unexplored, but fundamental phenomenon a {\it resonant reaction} in the field of diffusion-limited molecular reactions.

\begin{figure}[t]
        \includegraphics[width = .35 \textwidth]{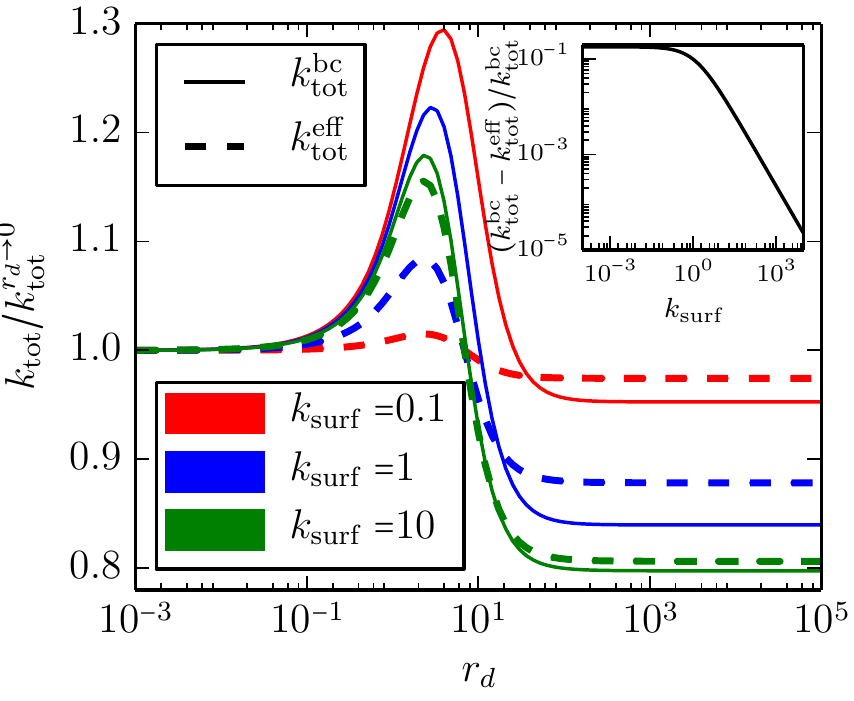}
        \caption{
        Comparison of the {\it total} reaction rate $k_{tot}$ of a non-perfect sink with a finite surface reaction rate $k_{\rm surf}$, 
        either caclulated by standard relation  eq.~\eqref{KS_nonideal} ($k_{\rm tot}^{\rm eff}$; dashed lines) or by the exact relation eq.~\eqref{nonideal_sink} ($k_{\rm tot}^{\rm bc}$; solid lines).      
        System parameters are $l = 5$, $g=1$,     $U_0 = 0$ and $U_1 = 3 ~k_{\rm B} T$. 
        The inset shows the relative difference between approaches~\eqref{nonideal_sink} and~\eqref{KS_nonideal} versus $k_{\rm surf}$  at maximum resonance. 
        \label{fig5}
        } 
    \end{figure}

We now turn to non-perfect sinks where the boundary condition is not fully adsorbing but a surface reaction with rate $k_{\rm surf}$ can take 
place according to eq.~(\ref{nonideal_sink}). In this case, for a {\it non-fluctuating} potential the total reaction rate $k_{\rm tot}$ is given {\it exactly} by the relationship eq.~(\ref{KS_nonideal}). However, it turns out that this standard additivity equation is not valid anymore in the case of fluctuating barriers, if the time scale of the fluctuations is not fast enough.  We show this by using our framework to calculate on one hand the diffusion rate
over the fluctuating barrier for a perfect sink, and then use eq.~\eqref{KS_nonideal} to combine it with $k_{\rm surf}$ to obtain the total rate $k_{\rm tot}^{\rm eff}$.  On the other hand, we directly calculate the total rate by using the same framework and the correct boundary conditions \eqref{nonideal_sink} for non-perfect sinks to obtain $k_{\rm tot}^{bc}$.  For non-fluctuating potentials, these two procedures lead exactly to the same result. 
A comparison for fluctuating barriers is shown in Fig.~\ref{fig5}.  Here, large relative discrepancies are observed when $k_{\rm surf}$ becomes comparable to the system scales {\it close to resonance}, i.e., $k_{\rm surf}\simeq 1$ and $r_{\rm d} \simeq 1$ (all in units of $R_{\rm s}$ and $D$), which continuously grow and eventually saturate for decreasing $k_{\rm surf}$ (see inset to Fig.~\ref{fig5}).  Note that the standard law eq.~(\ref{KS_nonideal}) is still valid in the fast limit ($r_{\rm d}\rightarrow 0$) while not in the slow limit ($r_{\rm d}\rightarrow \infty$) which can be analytically proven~\cite{supp}. Our analysis demonstrates that diffusion, barrier crossing, and surface reaction processes all dynamically interact and can not be decoupled in general, as assumed in eq.~(\ref{KS_nonideal}). Hence, care has to be taken in the interpretation of reaction rate processes in fluctuating environments. 

In summary, we have demonstrated the existence of the phenomenon of resonance reaction in diffusion-influenced reaction processes.
 For non-perfect sinks, we have also shown that the standard reciprocal additivity of diffusion and surface reaction rates is violated. 
 The fundamental findings derived here could be helpful to interpret  reaction rates in complex reaction systems~\cite{Szabo1982, agmon, srajer, zhou, greives, Setny2013, Mondal2013, kirkwood,mikael, polymertranslocation, wonkyu, Wu2012a, nanoreactor}, as well as for the control and optimization of association speeds in functional material design.  Although we have explored only symmetric fluctuation in a two-state model, our framework can deal with asymmetric switching rates between multiple states, greatly increasing the complexity of the problem due to the introduction of a full spectrum of  fluctuation time scales. 

\acknowledgments
The authors thank the Alexander von Humboldt (AvH) Foundation and the Foundation of German Industries (SDW) for financial support.  
J.D. acknowledges funding by the ERC (European Research Council) Consolidator Grant with project number 646659--NANOREACTOR.

%


\begin{thebibliography}{30}%
\makeatletter
\providecommand \@ifxundefined [1]{%
 \@ifx{#1\undefined}
}%
\providecommand \@ifnum [1]{%
 \ifnum #1\expandafter \@firstoftwo
 \else \expandafter \@secondoftwo
 \fi
}%
\providecommand \@ifx [1]{%
 \ifx #1\expandafter \@firstoftwo
 \else \expandafter \@secondoftwo
 \fi
}%
\providecommand \natexlab [1]{#1}%
\providecommand \enquote  [1]{``#1''}%
\providecommand \bibnamefont  [1]{#1}%
\providecommand \bibfnamefont [1]{#1}%
\providecommand \citenamefont [1]{#1}%
\providecommand \href@noop [0]{\@secondoftwo}%
\providecommand \href [0]{\begingroup \@sanitize@url \@href}%
\providecommand \@href[1]{\@@startlink{#1}\@@href}%
\providecommand \@@href[1]{\endgroup#1\@@endlink}%
\providecommand \@sanitize@url [0]{\catcode `\\12\catcode `\$12\catcode
  `\&12\catcode `\#12\catcode `\^12\catcode `\_12\catcode `\%12\relax}%
\providecommand \@@startlink[1]{}%
\providecommand \@@endlink[0]{}%
\providecommand \url  [0]{\begingroup\@sanitize@url \@url }%
\providecommand \@url [1]{\endgroup\@href {#1}{\urlprefix }}%
\providecommand \urlprefix  [0]{URL }%
\providecommand \Eprint [0]{\href }%
\providecommand \doibase [0]{http://dx.doi.org/}%
\providecommand \selectlanguage [0]{\@gobble}%
\providecommand \bibinfo  [0]{\@secondoftwo}%
\providecommand \bibfield  [0]{\@secondoftwo}%
\providecommand \translation [1]{[#1]}%
\providecommand \BibitemOpen [0]{}%
\providecommand \bibitemStop [0]{}%
\providecommand \bibitemNoStop [0]{.\EOS\space}%
\providecommand \EOS [0]{\spacefactor3000\relax}%
\providecommand \BibitemShut  [1]{\csname bibitem#1\endcsname}%
\let\auto@bib@innerbib\@empty
\bibitem [{\citenamefont {Calef}\ and\ \citenamefont
  {Deutch}(1983)}]{Calef1983}%
  \BibitemOpen
  \bibfield  {author} {\bibinfo {author} {\bibfnamefont {D.~F.}\ \bibnamefont
  {Calef}}\ and\ \bibinfo {author} {\bibfnamefont {J.~M.}\ \bibnamefont
  {Deutch}},\ }\href {\doibase 10.1146/annurev.pc.34.100183.002425} {\bibfield
  {journal} {\bibinfo  {journal} {Annu. Rev. Phys. Chem.}\ }\textbf {\bibinfo
  {volume} {34}},\ \bibinfo {pages} {493} (\bibinfo {year} {1983})}\BibitemShut
  {NoStop}%
\bibitem [{\citenamefont {Berg}\ and\ \citenamefont {von
  Hippel}(1985)}]{berg1985diffusion}%
  \BibitemOpen
  \bibfield  {author} {\bibinfo {author} {\bibfnamefont {O.~G.}\ \bibnamefont
  {Berg}}\ and\ \bibinfo {author} {\bibfnamefont {P.~H.}\ \bibnamefont {von
  Hippel}},\ }\href@noop {} {\bibfield  {journal} {\bibinfo  {journal} {Annu.
  Rev. Biophys. Biophys. Chem.}\ }\textbf {\bibinfo {volume} {14}},\ \bibinfo
  {pages} {131} (\bibinfo {year} {1985})}\BibitemShut {NoStop}%
\bibitem [{\citenamefont {Smoluchowski}(1917)}]{Smoluchowski1917a}%
  \BibitemOpen
  \bibfield  {author} {\bibinfo {author} {\bibfnamefont {M.}~\bibnamefont
  {von Smoluchowski}},\ }\href@noop {} {\bibfield  {journal} {\bibinfo  {journal}
  {Z. Phys. Chem.}\ }\textbf {\bibinfo {volume} {92}},\ \bibinfo {pages} {129}
  (\bibinfo {year} {1917})}\BibitemShut {NoStop}%
\bibitem [{\citenamefont {Debye}(1942)}]{Debye1942}%
  \BibitemOpen
  \bibfield  {author} {\bibinfo {author} {\bibfnamefont {P.}~\bibnamefont
  {Debye}},\ }\href@noop {} {\bibfield  {journal} {\bibinfo  {journal} {Trans.
  Electrochem. Soc.}\ }\textbf {\bibinfo {volume} {82}},\ \bibinfo {pages} {265}
  (\bibinfo {year} {1942})}\BibitemShut {NoStop}%
\bibitem [{\citenamefont {Kang}\ and\ \citenamefont {Redner}(1985)}]{kang}%
  \BibitemOpen
  \bibfield  {author} {\bibinfo {author} {\bibfnamefont {K.}~\bibnamefont
  {Kang}}\ and\ \bibinfo {author} {\bibfnamefont {S.}~\bibnamefont {Redner}},\
  }\href@noop {} {\bibfield  {journal} {\bibinfo  {journal} {Phys. Rev. A}\
  }\textbf {\bibinfo {volume} {32}},\ \bibinfo {pages} {435} (\bibinfo {year}
  {1985})}\BibitemShut {NoStop}%
\bibitem [{\citenamefont {Zwanzig}(1990)}]{zwanzig}%
  \BibitemOpen
  \bibfield  {author} {\bibinfo {author} {\bibfnamefont {R.}~\bibnamefont
  {Zwanzig}},\ }\href@noop {} {\bibfield  {journal} {\bibinfo  {journal} {Acc.
  Chem. Res.}\ }\textbf {\bibinfo {volume} {23}},\ \bibinfo {pages} {148}
  (\bibinfo {year} {1990})}\BibitemShut {NoStop}%
\bibitem [{\citenamefont {Szabo}\ \emph {et~al.}(1982)\citenamefont {Szabo},
  \citenamefont {Shoup}, \citenamefont {Northrup},\ and\ \citenamefont
  {McCammon}}]{Szabo1982}%
  \BibitemOpen
  \bibfield  {author} {\bibinfo {author} {\bibfnamefont {A.}~\bibnamefont
  {Szabo}}, \bibinfo {author} {\bibfnamefont {D.}~\bibnamefont {Shoup}},
  \bibinfo {author} {\bibfnamefont {S.~H.}\ \bibnamefont {Northrup}}, \ and\
  \bibinfo {author} {\bibfnamefont {J.~A.}\ \bibnamefont {McCammon}},\ }\href
  {\doibase 10.1063/1.444397} {\bibfield  {journal} {\bibinfo  {journal} {J.
  Chem. Phys.}\ }\textbf {\bibinfo {volume} {77}},\ \bibinfo {pages} {4484}
  (\bibinfo {year} {1982})}\BibitemShut {NoStop}%
\bibitem [{\citenamefont {Agmon}\ and\ \citenamefont {Hopfield}(1983)}]{agmon}%
  \BibitemOpen
  \bibfield  {author} {\bibinfo {author} {\bibfnamefont {N.}~\bibnamefont
  {Agmon}}\ and\ \bibinfo {author} {\bibfnamefont {J.~J.}\ \bibnamefont
  {Hopfield}},\ }\href@noop {} {\bibfield  {journal} {\bibinfo  {journal} {J.
  Chem. Phys.}\ }\textbf {\bibinfo {volume} {78}},\ \bibinfo {pages} {6947}
  (\bibinfo {year} {1983})}\BibitemShut {NoStop}%
\bibitem [{\citenamefont {Srajer}\ \emph {et~al.}(1988)\citenamefont {Srajer},
  \citenamefont {Reinisch},\ and\ \citenamefont {Champion}}]{srajer}%
  \BibitemOpen
  \bibfield  {author} {\bibinfo {author} {\bibfnamefont {V.}~\bibnamefont
  {Srajer}}, \bibinfo {author} {\bibfnamefont {L.}~\bibnamefont {Reinisch}}, \
  and\ \bibinfo {author} {\bibfnamefont {P.~M.}\ \bibnamefont {Champion}},\
  }\href@noop {} {\bibfield  {journal} {\bibinfo  {journal} {J. Am. Chem.
  Soc.}\ }\textbf {\bibinfo {volume} {110}},\ \bibinfo {pages} {6656} (\bibinfo
  {year} {1988})}\BibitemShut {NoStop}%
\bibitem [{\citenamefont {Zhou}\ \emph {et~al.}(1998)\citenamefont {Zhou},
  \citenamefont {Wlodek},\ and\ \citenamefont {McCammon}}]{zhou}%
  \BibitemOpen
  \bibfield  {author} {\bibinfo {author} {\bibfnamefont {H.-X.}\ \bibnamefont
  {Zhou}}, \bibinfo {author} {\bibfnamefont {S.~T.}\ \bibnamefont {Wlodek}}, \
  and\ \bibinfo {author} {\bibfnamefont {J.~A.}\ \bibnamefont {McCammon}},\
  }\href@noop {} {\bibfield  {journal} {\bibinfo  {journal} {Proc. Natl. Acad.
  Sci. (USA)}\ }\textbf {\bibinfo {volume} {95}},\ \bibinfo {pages} {9280}
  (\bibinfo {year} {1998})}\BibitemShut {NoStop}%
\bibitem [{\citenamefont {Greives}\ and\ \citenamefont {Zhou}(2014)}]{greives}%
  \BibitemOpen
  \bibfield  {author} {\bibinfo {author} {\bibfnamefont {N.}~\bibnamefont
  {Greives}}\ and\ \bibinfo {author} {\bibfnamefont {H.-X.}\ \bibnamefont
  {Zhou}},\ }\href@noop {} {\bibfield  {journal} {\bibinfo  {journal} {Proc.
  Natl. Acad. Sci. USA.}\ }\textbf {\bibinfo {volume} {111}},\ \bibinfo {pages}
  {10197} (\bibinfo {year} {2014})}\BibitemShut {NoStop}%
\bibitem [{\citenamefont {Setny}\ \emph {et~al.}(2013)\citenamefont {Setny},
  \citenamefont {Baron}, \citenamefont {{M. Kekenes-Huskey}}, \citenamefont
  {McCammon},\ and\ \citenamefont {Dzubiella}}]{Setny2013}%
  \BibitemOpen
  \bibfield  {author} {\bibinfo {author} {\bibfnamefont {P.}~\bibnamefont
  {Setny}}, \bibinfo {author} {\bibfnamefont {R.}~\bibnamefont {Baron}},
  \bibinfo {author} {\bibfnamefont {P.}~\bibnamefont {{M. Kekenes-Huskey}}},
  \bibinfo {author} {\bibfnamefont {J.~A.}\ \bibnamefont {McCammon}}, \ and\
  \bibinfo {author} {\bibfnamefont {J.}~\bibnamefont {Dzubiella}},\ }\href
  {\doibase 10.1073/pnas.1221231110} {\bibfield  {journal} {\bibinfo  {journal}
  {Proc. Natl. Acad. Sci. U. S. A.}\ }\textbf {\bibinfo {volume} {110}},\
  \bibinfo {pages} {1197} (\bibinfo {year} {2013})}\BibitemShut {NoStop}%
\bibitem [{\citenamefont {Mondal}\ \emph {et~al.}(2013)\citenamefont {Mondal},
  \citenamefont {Morrone},\ and\ \citenamefont {Berne}}]{Mondal2013}%
  \BibitemOpen
  \bibfield  {author} {\bibinfo {author} {\bibfnamefont {J.}~\bibnamefont
  {Mondal}}, \bibinfo {author} {\bibfnamefont {J.}~\bibnamefont {Morrone}}, \
  and\ \bibinfo {author} {\bibfnamefont {B.~J.}\ \bibnamefont {Berne}},\ }\href
  {\doibase 10.1073/pnas.1312529110} {\bibfield  {journal} {\bibinfo  {journal}
  {Proc. Natl. Acad. Sci. U. S. A.}\ }\textbf {\bibinfo {volume} {110}},\
  \bibinfo {pages} {13277} (\bibinfo {year} {2013})}\BibitemShut {NoStop}%
\bibitem [{\citenamefont {Kirkwook}\ and\ \citenamefont
  {Shumaker}(1952)}]{kirkwood}%
  \BibitemOpen
  \bibfield  {author} {\bibinfo {author} {\bibfnamefont {J.~G.}\ \bibnamefont
  {Kirkwood}}\ and\ \bibinfo {author} {\bibfnamefont {J.~B.}\ \bibnamefont
  {Shumaker}},\ }\href@noop {} {\bibfield  {journal} {\bibinfo  {journal}
  {Proc. Natl. Acad. Sci. USA.}\ }\textbf {\bibinfo {volume} {38}},\ \bibinfo
  {pages} {863} (\bibinfo {year} {1952})}\BibitemShut {NoStop}%
\bibitem [{\citenamefont {Lund}\ and\ \citenamefont
  {J{\"o}nsson}(2013)}]{mikael}%
  \BibitemOpen
  \bibfield  {author} {\bibinfo {author} {\bibfnamefont {M.}~\bibnamefont
  {Lund}}\ and\ \bibinfo {author} {\bibfnamefont {B.}~\bibnamefont
  {J{\"o}nsson}},\ }\href@noop {} {\bibfield  {journal} {\bibinfo  {journal}
  {Quart. Rev. Biophys.}\ }\textbf {\bibinfo {volume} {46}},\ \bibinfo {pages}
  {265} (\bibinfo {year} {2013})}\BibitemShut {NoStop}%
\bibitem [{\citenamefont {Pizzolato}\ \emph {et~al.}(2010)\citenamefont
  {Pizzolato}, \citenamefont {Fiasconaro}, \citenamefont {Adorno},\ and\
  \citenamefont {Spagnolo}}]{polymertranslocation}%
  \BibitemOpen
  \bibfield  {author} {\bibinfo {author} {\bibfnamefont {N.}~\bibnamefont
  {Pizzolato}}, \bibinfo {author} {\bibfnamefont {A.}~\bibnamefont
  {Fiasconaro}}, \bibinfo {author} {\bibfnamefont {D.~P.}\ \bibnamefont
  {Adorno}}, \ and\ \bibinfo {author} {\bibfnamefont {B.}~\bibnamefont
  {Spagnolo}},\ }\href@noop {} {\bibfield  {journal} {\bibinfo  {journal}
  {Physical Biology}\ }\textbf {\bibinfo {volume} {7}},\ \bibinfo {pages}
  {034001} (\bibinfo {year} {2010})}\BibitemShut {NoStop}%
\bibitem [{\citenamefont {Kim}\ \emph {et~al.}(2012)\citenamefont {Kim},
  \citenamefont {Hyeon},\ and\ \citenamefont {Sung}}]{wonkyu}%
  \BibitemOpen
  \bibfield  {author} {\bibinfo {author} {\bibfnamefont {W.~K.}\ \bibnamefont
  {Kim}}, \bibinfo {author} {\bibfnamefont {C.}~\bibnamefont {Hyeon}}, \ and\
  \bibinfo {author} {\bibfnamefont {W.}~\bibnamefont {Sung}},\ }\href@noop {}
  {\bibfield  {journal} {\bibinfo  {journal} {Proc. Natl. Acad. Sci.}\ }\textbf
  {\bibinfo {volume} {109}},\ \bibinfo {pages} {14410} (\bibinfo {year}
  {2012})}\BibitemShut {NoStop}%
\bibitem [{\citenamefont {Wu}\ \emph {et~al.}(2012)\citenamefont {Wu},
  \citenamefont {Dzubiella}, \citenamefont {Kaiser}, \citenamefont {Drechsler},
  \citenamefont {Guo}, \citenamefont {Ballauff},\ and\ \citenamefont
  {Lu}}]{Wu2012a}%
  \BibitemOpen
  \bibfield  {author} {\bibinfo {author} {\bibfnamefont {S.}~\bibnamefont
  {Wu}}, \bibinfo {author} {\bibfnamefont {J.}~\bibnamefont {Dzubiella}},
  \bibinfo {author} {\bibfnamefont {J.}~\bibnamefont {Kaiser}}, \bibinfo
  {author} {\bibfnamefont {M.}~\bibnamefont {Drechsler}}, \bibinfo {author}
  {\bibfnamefont {X.}~\bibnamefont {Guo}}, \bibinfo {author} {\bibfnamefont
  {M.}~\bibnamefont {Ballauff}}, \ and\ \bibinfo {author} {\bibfnamefont
  {Y.}~\bibnamefont {Lu}},\ }\href {\doibase 10.1002/anie.201106515} {\bibfield
   {journal} {\bibinfo  {journal} {Angew. Chem. Int. Ed. Engl.}\ }\textbf
  {\bibinfo {volume} {51}},\ \bibinfo {pages} {2229} (\bibinfo {year}
  {2012})}\BibitemShut {NoStop}%
\bibitem [{\citenamefont {Angioletti-Uberti}\ \emph {et~al.}(2015)\citenamefont
  {Angioletti-Uberti}, \citenamefont {Lu}, \citenamefont {Ballauff},\ and\
  \citenamefont {Dzubiella}}]{nanoreactor}%
  \BibitemOpen
  \bibfield  {author} {\bibinfo {author} {\bibfnamefont {S.}~\bibnamefont
  {Angioletti-Uberti}}, \bibinfo {author} {\bibfnamefont {Y.}~\bibnamefont
  {Lu}}, \bibinfo {author} {\bibfnamefont {M.}~\bibnamefont {Ballauff}}, \ and\
  \bibinfo {author} {\bibfnamefont {J.}~\bibnamefont {Dzubiella}},\ }\href@noop
  {} {\bibfield  {journal} {\bibinfo  {journal} {J. Phys. Chem. C}\ }\textbf
  {\bibinfo {volume} {119}},\ \bibinfo {pages} {15723} (\bibinfo {year}
  {2015})}\BibitemShut {NoStop}%
\bibitem [{\citenamefont {Doering}\ and\ \citenamefont
  {Gadoua}(1992)}]{Doering1992}%
  \BibitemOpen
  \bibfield  {author} {\bibinfo {author} {\bibfnamefont {C.~R.}~\bibnamefont
  {Doering}}\ and\ \bibinfo {author} {\bibfnamefont {J.~C.}~\bibnamefont
  {Gadoua}},\ }\href {http://prl.aps.org/abstract/PRL/v69/i16/p2318\_1}
  {\bibfield  {journal} {\bibinfo  {journal} {Phys. Rev. Lett.}\ }\textbf
  {\bibinfo {volume} {69}},\ \bibinfo {pages} {2318} (\bibinfo {year}
  {1992})}\BibitemShut {NoStop}%
\bibitem [{\citenamefont {Z\"{u}rcher}\ and\ \citenamefont
  {Doering}(1993)}]{Zurcher1993}%
  \BibitemOpen
  \bibfield  {author} {\bibinfo {author} {\bibfnamefont {U.}~\bibnamefont
  {Z\"{u}rcher}}\ and\ \bibinfo {author} {\bibfnamefont {C.~R.}~\bibnamefont
  {Doering}},\ }\href {http://pre.aps.org/abstract/PRE/v47/i6/p3862\_1}
  {\bibfield  {journal} {\bibinfo  {journal} {Phys. Rev. E}\ }\textbf {\bibinfo
  {volume} {47}},\ \bibinfo {pages} {3862} (\bibinfo {year}
  {1993})}\BibitemShut {NoStop}%
\bibitem [{\citenamefont {Pechukas}\ and\ \citenamefont
  {H\"{a}nggi}(1994)}]{Pechukas1994}%
  \BibitemOpen
  \bibfield  {author} {\bibinfo {author} {\bibfnamefont {P.}~\bibnamefont
  {Pechukas}}\ and\ \bibinfo {author} {\bibfnamefont {P.}~\bibnamefont
  {H\"{a}nggi}},\ }\href {http://prl.aps.org/abstract/PRL/v73/i20/p2772\_1}
  {\bibfield  {journal} {\bibinfo  {journal} {Phys. Rev. Lett.}\ }\textbf
  {\bibinfo {volume} {73}},\ \bibinfo {pages} {2772} (\bibinfo {year}
  {1994})}\BibitemShut {NoStop}%
\bibitem [{\citenamefont {Reimann}(1995{\natexlab{a}})}]{Reimann1995}%
  \BibitemOpen
  \bibfield  {author} {\bibinfo {author} {\bibfnamefont {P.}~\bibnamefont
  {Reimann}},\ }\href {http://pre.aps.org/abstract/PRE/v52/i2/p1579\_1}
  {\bibfield  {journal} {\bibinfo  {journal} {Phys. Rev. E}\ }\textbf {\bibinfo
  {volume} {52}},\ \bibinfo {pages} {1579} (\bibinfo {year} {1995}{\natexlab{a}})}\BibitemShut {NoStop}%
\bibitem [{\citenamefont {Reimann}(1995{\natexlab{b}})}]{Reimann1995a}%
  \BibitemOpen
  \bibfield  {author} {\bibinfo {author} {\bibfnamefont {P.}~\bibnamefont
  {Reimann}},\ }\href {http://prl.aps.org/abstract/PRL/v74/i23/p4576\_1}
  {\bibfield  {journal} {\bibinfo  {journal} {Phys. Rev. Lett.}\ }\textbf
  {\bibinfo {volume} {74}},\ \bibinfo {pages} {4576} (\bibinfo {year}
  {1995}{\natexlab{b}})}\BibitemShut {NoStop}%
\bibitem [{\citenamefont {Reimann}\ and\ \citenamefont
  {H\"{a}nggi}(1997)}]{Reimann1997}%
  \BibitemOpen
  \bibfield  {author} {\bibinfo {author} {\bibfnamefont {P.}~\bibnamefont
  {Reimann}}\ and\ \bibinfo {author} {\bibfnamefont {P.}~\bibnamefont
  {H\"{a}nggi}},\ }\href@noop {} {\emph {\bibinfo {title} {{Surmounting
  fluctuating barriers: basic concepts and results}}}},\ edited by\ \bibinfo
  {editor} {\bibfnamefont {L.}~\bibnamefont {Schimansky-Geier}}\ and\ \bibinfo
  {editor} {\bibfnamefont {T.}~\bibnamefont {P\"{o}schel}},\ Vol.\ \bibinfo
  {volume} {484}\ (\bibinfo  {publisher} {Springer Verlag, Berlin},\ \bibinfo
  {year} {1997})\ pp.\ \bibinfo {pages} {127--139}\BibitemShut {NoStop}%
\bibitem [{\citenamefont {Mantegna}\ and\ \citenamefont
  {Spagnolo}(2000)}]{mantegna}%
  \BibitemOpen
  \bibfield  {author} {\bibinfo {author} {\bibfnamefont {R.~N.}\ \bibnamefont
  {Mantegna}}\ and\ \bibinfo {author} {\bibfnamefont {B.}~\bibnamefont
  {Spagnolo}},\ }\href@noop {} {\bibfield  {journal} {\bibinfo  {journal}
  {Phys. Rev. Lett.}\ }\textbf {\bibinfo {volume} {84}},\ \bibinfo {pages}
  {3025} (\bibinfo {year} {2000})}\BibitemShut {NoStop}%
\bibitem [{\citenamefont {Schmitt}\ and\ \citenamefont
  {Schmitt}(2004)}]{schmitt}%
  \BibitemOpen
  \bibfield  {author} {\bibinfo {author} {\bibfnamefont {C.}\ \bibnamefont
  {Schmitt}},\ \bibinfo {author} {\bibfnamefont {B.}\ \bibnamefont
  {Dybiec}},\ \bibinfo {author} {\bibfnamefont {P.}\ \bibnamefont
  {H\"anggi}}\ and\ \bibinfo {author} {\bibfnamefont {C.}~\bibnamefont
  {Bechinger}},\ }\href@noop {} {\bibfield  {journal} {\bibinfo  {journal}
  {Europhys. Lett.}\ }\textbf {\bibinfo {volume} {74}},\ \bibinfo {pages}
  {937} (\bibinfo {year} {2006})}\BibitemShut {NoStop}%
\bibitem [{\citenamefont {Oppenheim}\ \emph {et~al.}(1977)\citenamefont
  {Oppenheim}, \citenamefont {Shuler},\ and\ \citenamefont
  {Weiss}}]{Oppenheim1977}%
  \BibitemOpen
  \bibfield  {author} {\bibinfo {author} {\bibfnamefont {I.}~\bibnamefont
  {Oppenheim}}, \bibinfo {author} {\bibfnamefont {K.~E.}\ \bibnamefont
  {Shuler}}, \ and\ \bibinfo {author} {\bibfnamefont {G.~H.}\ \bibnamefont
  {Weiss}},\ }\href@noop {} {\emph {\bibinfo {title} {{Stochastic Processes in
  Chemical Physics: The Master Equation}}}},\ \bibinfo {edition} {1st}\ ed.\
  (\bibinfo  {publisher} {The MIT Press},\ \bibinfo {year} {1977})\ pp.\
  \bibinfo {pages} {67----85}\BibitemShut {NoStop}%
\bibitem [{\citenamefont {{Van Kampen}}(1992)}]{VanKampen1992}%
  \BibitemOpen
  \bibfield  {author} {\bibinfo {author} {\bibfnamefont {N.}~\bibnamefont {{Van
  Kampen}}},\ }\href@noop {} {\emph {\bibinfo {title} {{Stochastic Processes in
  Physics and Chemistry}}}},\ \bibinfo {edition} {2nd}\ ed.\ (\bibinfo
  {publisher} {North-Holland Personal Library},\ \bibinfo {year} {1992})\ pp.\
  \bibinfo {pages} {104----108}\BibitemShut {NoStop}%
\bibitem [{\citenamefont {{supp}}(1992)}]{supp}%
  \BibitemOpen
  See supplemental material at [URL will be inserted by AIP] for technical details of the analytical and simulation methods as well as the analysis
  \BibitemShut {NoStop}%
\bibitem [{\citenamefont {{Wolfram Research, Inc.}}(2012)}]{mathematica}%
  \BibitemOpen
  \bibfield  {author} {\bibinfo {author} {\bibnamefont {{Wolfram Research,
  Inc.}}},\ }\href@noop {} {\emph {\bibinfo {title} {Mathematica, Version
  9.0}}}\ (\bibinfo  {publisher} {Wolfram Research, Inc.},\ \bibinfo {address}
  {Champaign, Illinois},\ \bibinfo {year} {2012})\BibitemShut {NoStop}%
\bibitem [{\citenamefont {andy}\ and\ \citenamefont
  {andy}(1983)}]{andy}%
  \BibitemOpen
  \bibfield  {author} {\bibinfo {author} {\bibfnamefont {S.~H.}~\bibnamefont
  {Northrup}},\  \bibinfo {author} {\bibfnamefont {S.~A.}~\bibnamefont
  {Allison}},\ and\ \bibinfo {author} {\bibfnamefont {J.~A.}\ \bibnamefont
  {McCammon}},\ }\href {\doibase 10.1063/1.1887165} {\bibfield  {journal}
  {\bibinfo  {journal} {J. Chem. Phys.}\ }\textbf {\bibinfo {volume} {80}},\
  \bibinfo {pages} {1517} (\bibinfo {year} {1983})}\BibitemShut {NoStop}%
\bibitem [{\citenamefont {Dzubiella}\ and\ \citenamefont
  {McCammon}(2005)}]{Dzubiella2005}%
  \BibitemOpen
  \bibfield  {author} {\bibinfo {author} {\bibfnamefont {J.}~\bibnamefont
  {Dzubiella}}\ and\ \bibinfo {author} {\bibfnamefont {J.~A.}\ \bibnamefont
  {McCammon}},\ }\href {\doibase 10.1063/1.1887165} {\bibfield  {journal}
  {\bibinfo  {journal} {J. Chem. Phys.}\ }\textbf {\bibinfo {volume} {122}},\
  \bibinfo {pages} {184902} (\bibinfo {year} {2005})}\BibitemShut {NoStop}%
\end{thebibliography}

\end{document}